\documentclass[preprint]{aastex62}

\received{3 October 2020}
\revised{16 November 2020}
\accepted{24 November 2020}

\submitjournal{ApJ}

\shorttitle{In situ evidence of ion acceleration between consecutive reconnection jet fronts}
\shortauthors{Catapano et al.}
\begin{document}

\title{In situ evidence of ion acceleration between consecutive reconnection jet fronts}

\correspondingauthor{Filomena Catapano}
\email{menacata3@gmail.com}

\author[0000-0002-2802-3920]{Filomena Catapano}
\affil{Laboratoire de Physique des Plasmas, CNRS/Ecole Polytechnique/Sorbonne Universit\'e, Universit\'e Paris Sud, Observatoire de Paris, Paris, France}
\affil{Dipartimento di Fisica, Universit\`a della Calabria, Rende, Italy}

\author{Alessandro Retin\`o}
\affil{Laboratoire de Physique des Plasmas, CNRS/Ecole Polytechnique/Sorbonne Universit\'e, Universit\'e Paris Sud, Observatoire de Paris, Paris, France}

\author{Gaetano Zimbardo}
\affil{Dipartimento di Fisica, Universit\`a della Calabria, Rende, Italy}

\author{Alexandra Alexandrova}
\affil{Laboratoire de Physique des Plasmas, CNRS/Ecole Polytechnique/Sorbonne Universit\'e, Universit\'e Paris Sud, Observatoire de Paris, Paris, France}

\author{Ian J. Cohen}
\affil{Applied Physics Laboratory, The Johns Hopkins University, Laurel, Maryland, US}

\author{Drew L. Turner}
\affil{Applied Physics Laboratory, The Johns Hopkins University, Laurel, Maryland, US}

\author{Olivier Le Contel}
\affil{Laboratoire de Physique des Plasmas, CNRS/Ecole Polytechnique/Sorbonne Universit\'e, Universit\'e Paris Sud, Observatoire de Paris, Paris, France}

\author{Giulia Cozzani}
\affil{Swedish Institute of Space Physics, Uppsala, Sweden}

\author{Silvia Perri}
\affil{Dipartimento di Fisica, Universit\`a della Calabria, Rende, Italy}

\author{Antonella Greco}
\affil{Dipartimento di Fisica, Universit\`a della Calabria, Rende, Italy}

\author{Hugo Breuillard}
\affil{Laboratoire de Physique des Plasmas, CNRS/Ecole Polytechnique/Sorbonne Universit\'e, Universit\'e Paris Sud, Observatoire de Paris, Paris, France}
\affil{Laboratoire de Physique et de Chimie de l'Environnement et de l'Espace/CNRS, Orleans, France}

\author{Dominique Delcourt}
\affil{Laboratoire de Physique et de Chimie de l'Environnement et de l'Espace/CNRS, Orleans, France}

\author{Laurent Mirioni}
\affil{Laboratoire de Physique des Plasmas, CNRS/Ecole Polytechnique/Sorbonne Universit\'e, Universit\'e Paris Sud, Observatoire de Paris, Paris, France}

\author{Yuri Khotyaintsev}
\affil{Swedish Institute of Space Physics, Uppsala, Sweden}

\author{Andris Vaivads}
\affil{Swedish Institute of Space Physics, Uppsala, Sweden}

\author{Barbara L. Giles}
\affil{Goddard Space Flight Center, Greenbelt, Maryland, US}

\author{Barry H. Mauk}
\affil{Applied Physics Laboratory, The Johns Hopkins University, Laurel, Maryland, US}

\author{Stephen A. Fuselier}
\affil{Southwest Research Institute}
\affil{University of Texas, San Antonio, Texas, US}

\author{Roy B. Torbert}
\affil{University of New Hampshire, Durham, New Hampshire, US}

\author{Christopher T. Russell}
\affil{Department of Earth and Space Science, University of California, Los Angles, California, US}

\author{Per A. Lindqvist}
\affil{Royal Institute of Technology, Stockholm, Sweden}

\author{Robert E. Ergun}
\affil{LASP, University of Colorado, Boulder, Colorado, US}

\author{Thomas Moore}
\affil{Goddard Space Flight Center, Greenbelt, Maryland, US}

\author{James L. Burch}
\affil{Southwest Research Institute}

\begin{abstract}

Processes driven by unsteady reconnection can efficiently accelerate particles in many astrophysical plasmas. An example are the reconnection jet fronts in an outflow region. We present evidence of suprathermal ion acceleration between two consecutive reconnection jet fronts observed by the Magnetospheric Multiscale mission in the terrestrial magnetotail. An earthward propagating jet is approached by a second faster jet. Between the jets, the thermal ions are mostly perpendicular to magnetic field,  are trapped and are gradually accelerated in the parallel direction up to $150$ keV. Observations suggest that ions are predominantly accelerated by a Fermi-like mechanism in the contracting magnetic bottle formed between the two jet fronts. The ion acceleration mechanism is presumably efficient in other environments where jet fronts produced by variable rates of reconnection are common and where the interaction of multiple jet fronts can also develop a turbulent environment, e.g. in stellar and solar eruptions.

\end{abstract}


\section{Introduction} \label{sec:intro}

Understanding the conversion of magnetic energy into acceleration of charged particles is a long-standing problem in astrophysics. The fundamental process of magnetic reconnection transfers a significant fraction of the stored magnetic energy into plasma heating and acceleration of particles \citep{Priest2000}. 
Reconnection is known to play a role in particle acceleration in solar flares \citep{Emslie2004,Krucker2007,Lin2011}, planetary magnetospheres \citep{Slavin2010,Oieroset2011,Imada2015,Lu2020}, and more distant astrophysical environments \citep{Kronberg2004,Tavani2011,Striani2011,Cerutti2013}. 
Both numerical simulations \citep{Pritchett2006,aa2011,Birn2012,Dahlin2014} and in situ observations \citep{Retino2008,Fu2013GRL,Fu2013} clearly indicate that most of the acceleration occurs away from the reconnection site through processes induced by reconnection. An important example is the acceleration occurring around reconnection jet fronts. These fronts are the boundaries separating hot, fast plasma jets from colder ambient plasma at rest \citep{Yuri2011,Fu2013GRL}. Observations \citep{Fu2013} and simulations \citep{Sitnov2009} suggest that multiple jet fronts are signatures of unsteady reconnection. The mechanisms of energetic particle generation at plasma jets have been studied extensively using numerical simulations \citep{Wu2012,Birn2014,Greco15,Catapano17,Ukho2018} and spacecraft observations \citep{Sergeev2009,Zhou2010,aa2011,Gabrielse2014}, but these mechanisms are not fully understood. Also, the interaction of the jet with the ambient plasma/obstacles can become unstable and eventually result in the formation of turbulent structures that can contribute to particle acceleration \citep{Drake2006,Retino2007,Retino2008,Greco2010,Ergun2020}.

An intriguing model of particle acceleration at jets is that of collapsing magnetic traps forming in the reconnection outflow region during solar eruptions. In this model, the plasma jet, confined on the newly reconnected field lines, moves towards the solar surface until it encounters the underlying magnetic loop \citep{Somov97}. As a result of its motion, the length of the magnetic trap decreases. This decrease accelerates the trapped particles via a first-order Fermi process related to the conservation of the second adiabatic invariant, $J=\oint p_{\parallel} dl$, where $p_{\parallel}$ is the particle momentum along the magnetic field direction, and $l$ is the particle path. The particle kinetic energy increases in the magnetic trap until particles fall into the loss cone \citep{Giuliani2005,Zharkova2011,Somov2013,Birn2017}. As a result of the jet braking, particles in such magnetic field configurations may be trapped not only between the loop's foot points, but also between secondary mirroring points along the flux tube \citep{Borissov16}. 
A similar first-order Fermi mechanism, combined with a betatron mechanism, is also invoked in planetary magnetotails \citep{Birn2012,Birn2014} where multiple jet fronts originate from unsteady reconnection \citep{Daughton2006,Eastwood2009,AlexandrovaA2012}.
 
Reconnection is common for most of the spatial and temporal scales in astrophysics, therefore understanding acceleration mechanisms driven by reconnection has crucial implications for all astrophysical plasmas. Electron acceleration at jet fronts has been observed recently in the terrestrial magnetotail \citep{Fu2013GRL,Fu2013,Turner2016,Hugo2017,Torbert2018}.
In situ observations for ion acceleration in such a scenario are still lacking. 

The recent Magnetospheric Multiscale (MMS) mission \citep{Burch2016}, with its unprecedented high-resolution plasma measurements, allows us to investigate ion acceleration in the Earth's magnetosphere in detail, studying the evolution of the ion distribution functions and ion moments with a very high temporal cadence.
Here we present MMS observations of efficient ion acceleration between two consecutive reconnection jet fronts in the terrestrial magnetotail. We show that according to the ion energy and pitch-angle distributions, the inter-jets region acts as a contracting magnetic bottle, which accelerates the ions up to several times their initial thermal energy through a Fermi-like mechanism.

\section{Event Overview}

\begin{figure}
        \centering
        \includegraphics[width=0.99\textwidth]{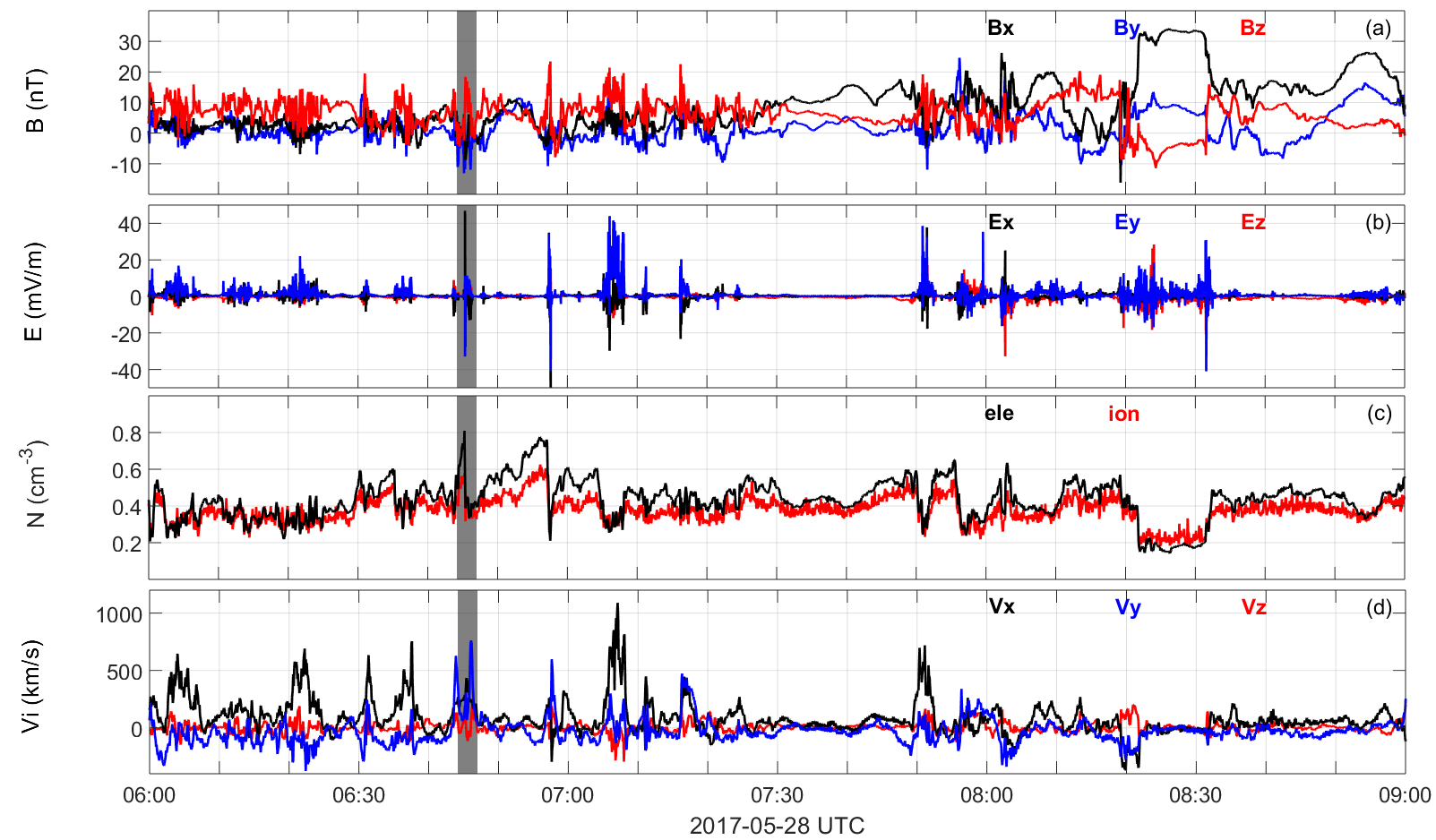}
        \caption{MMS observations on 28 May 2017 between 06:00 and 09:00 UT. Panels show the (a) magnetic field components ($x$, $y$, and $z$ represented with black, blue and red lines respectively) , (b) electric field components, (c) ion and electron density (red and black lines), (d) ion velocity components. The shadowed region identify the event studied in the paper.}\label{Fig1}
    \end{figure}

On 28 May 2017 between 06:00 and 09:00 UT, the four MMS satellites were in the Earth's magnetotail in a tetrahedral configuration with an inter-spacecraft separation of 20 km. The Dst (disturbance storm time) index was around $- 120$ nT indicating a moderate geomagnetic activity \citep{Dst}. Indeed, the magnetotail was very perturbed and MMS observed numerous earthward jets, as shown in Figure \ref{Fig1}. In panels (a), (b) and (d), the three components of the magnetic and electric fields measured by the Fluxgate Magnetometers (FGM) \citep{Russell2016} and the Electric field Double Probes (EDP) \citep{Torbert2016,Ergun2016,Lindqvist2016}, and ion velocity measured by FPI-DIS instrument \citep{Pollock2016}, are respectively shown. While, panel (c) shows the ion and electron densities measured by FPI instrument. From 06:00 up to almost 07:30 UT, the $B_x$ component of the electric field is almost close to zero, indicating that the spacecraft are close to the center of the current sheet all the time. After 07:30 UT, we observe an increase of the magnetic $B_x$ component up to 20 nT, indicating that the spacecraft are in a region closer to the magnetic lobes. Also, as shown in panel (d), the ion velocity fluctuates strongly during this time interval. The time period studied in details in this study is the one shaded in Figure \ref{Fig1}. It is characterized by very strong electric field $E_y$ and $E_x$ components (panel (b)), high density gradient (panel (c)), and strong ion velocity gradient (panel (d)). The gradients in the ion velocity, together with the density gradient, the strong electric field, and the strong $B_z$ variation (panel (a)), are signature of reconnection jet fronts \citep{Runov2009}. Furthermore, the strong variation of $V_y$ is typically observed in correspondence of multiple jet fronts \citep{Zhou2009}. We decided to focus our analysis on the event observed between 06:44:40 and 06:46:40 UT which is characterised by a dominant $V_y$ component of the ion velocity and other signatures indicative of effective ion acceleration processes that will be discussed in the following. 
 
\begin{figure}
        \centering
        \includegraphics[width=18cm, height=10cm]{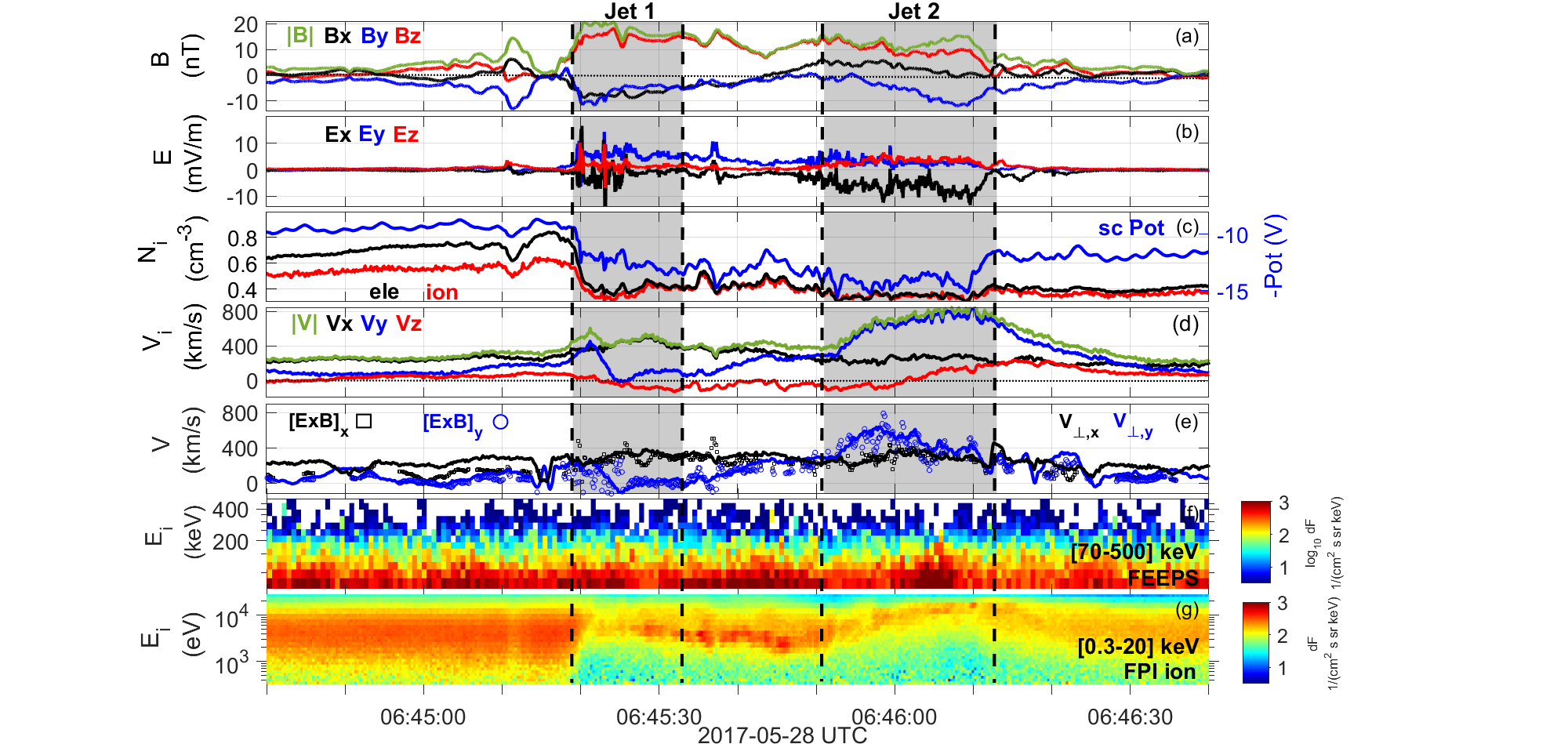}
        \caption{MMS observations on 28 May 2017 between 06:44:40 and 06:46:40 UT. Panels show the (a) magnetic field components ($x$, $y$, and $z$ represented with black, blue and red lines respectively) together with $|\textbf{B}|$ profile (green line), (b) electric field components, (c) ion and electron density (red and black lines) together with the spacecraft potential (blue line), (d) ion velocity components together with $|V|$ profile, (e) perpendicular ion velocity components (solid lines) together with $\textbf{E}\times\textbf{B}/B^2$ along $x$ and $y$ (squares and circles, profiles smoothed on 0.9 s), (f) ion energy spectrogram between [70-500] keV and (g) [0.3-20] keV. The two pairs of vertical dashed lines identify the passage of the two consecutive plasma jet fronts. All the data are in the GSM coordinate system and in burst mode.}\label{Fig2}
    \end{figure}

 Figure \ref{Fig2} is an overview of the selected event. MMS Spacecraft were at $\sim (-20, -12, -2)$ Earth radii ($R_E$) in the Geocentric Solar Magnetospheric (GSM) coordinate system. Since the spacecraft separation was much smaller than the typical proton gyroradius in the plasmasheet, that is $\rho_p \sim 500$ km, the four-probe averaged data are used. Panels \ref{Fig2}(a) and (b) show profiles of the magnetic and electric field components. At around 06:44:40 UT, all magnetic field components are small, indicating that the spacecraft were near the center of the current sheet (CS). At 06:45:29 UT (first dashed vertical line in Figure \ref{Fig2}), a sharp enhancement in the $B_z \sim |\textbf{B}|$ profile is observed, with the magnetic field increasing from zero (horizontal dashed line) to $\sim$16 nT. This enhancement is preceded by a magnetic dip, that is a typical jet front signature \citep{Runov2011}. The enhanced $B_z$ region is accompanied by intense electric field fluctuations, $E_y$ reaching $\sim$ 20 mV/m. The $B_x$ magnetic field component is negative, indicating that the spacecraft are south of the CS. After 06:45:33 UT (second dashed vertical line in Figure \ref{Fig2}) both magnetic and electric fields start to substantially decrease and at 06:45:43 UT $B_z \sim |\textbf{B}|$ has a local minimum and $B_x$ becomes positive, indicating that the spacecraft crossed the CS. Around 06:45:50 UT, $B_z \sim |\textbf{B}|$ and the electric field fluctuations become strong again (third dashed vertical line in Figure \ref{Fig2}). This suggests that MMS observed a second reconnection jet front. 
   
\begin{figure}
        \centering
        \includegraphics[width=0.8\textwidth]{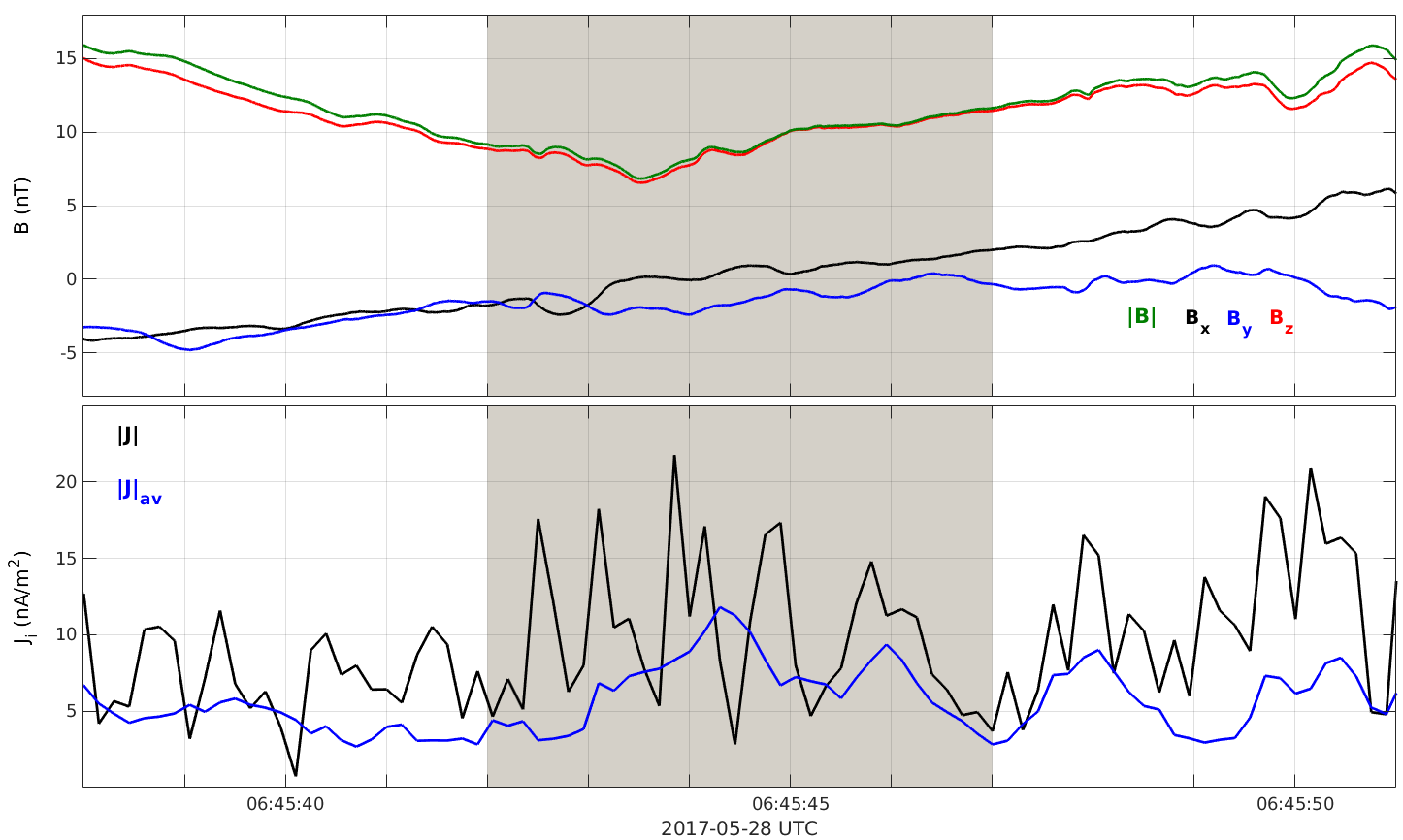}
        \caption{Zoom of the current sheet crossing in the region between the two reconnection jets. Upper panel shows the three magnetic field components ($x$, $y$ and $z$ represented with black, blue and red lines respectively) together with $|\textbf{B}|$ (green line). Bottom panel shows the ion current density $|\textbf{J}|$ (black line), and its average at 0.6 sec (blue line). }\label{Fig3}
    \end{figure}

  A detailed analysis of the CS crossing between the two reconnection jets is reported in Figure \ref{Fig3}. The three magnetic field components together with $|\textbf{B}|$ are shown in the upper panel. We observe $B_x$ varying between -5 and 5 nT and changing sign in correspondence with the $|\textbf{B}|$ minimum. In the second panel the ion current density $|\textbf{J}|$ (black line) is plotted. The current density was evaluated by using the plasma moments from FPI instrument, which are over-plotted at 0.6 sec resolution (blue line). We observe a current density peak (see blue line in the shaded region) in correspondence with the local minimum of $|\textbf{B}|$, where $B_x \sim 0$ nT. These profiles are typical signatures of current sheet crossings \citep{Runov2006}. 
   
Following the hypothesis of MMS observing two reconnection jet fronts, it is possible to distinguish three regions: the Jet 1 region, observed between 06:45:29-06:45:34 UT, the region between two jets 06:45:34-06:45:50 UT, and the Jet 2, observed between 06:45:50- 06:46:13 UT. The regions Jet 1 and Jet 2, shaded in Figure \ref{Fig2}, are characterized by the strong $E_y$ fluctuations and large $B_z$. The region in between the jets is characterized by the weaker electric field fluctuations and a local $|\textbf{B}|$ minimum.  
   
Figure \ref{Fig2}(f) shows the ion energy spectrogram detected by the Fly's Eye Energetic Particle Spectrometer (FEEPS) \citep{Mauk2016}, measuring energetic ions in the range [70-500] keV with a resolution of 300 ms. Figure \ref{Fig2}(g) shows the ion energy spectrogram detected by the FPI-DIS instrument, measuring ions in the range 30 eV-30 keV with a time resolution of 150 ms. Figure \ref{Fig2}(c) shows the ion density from FPI-DIS together with the electron density from FPI-DES and the negative of the spacecraft potential, which is a proxy of the plasma density (see \cite{Andriopoulou2018} and reference therein). At the beginning and ending of the event we observe typical ambient plasma sheet with an ion density of $\sim$ 0.5 cm$^{-3}$. We observe density gradients in correspondence with the two sharp $B_z$ enhancements (first and last dashed lines in Figure \ref{Fig2}) and the density decreases to 50\% of the value observed in the ambient plasma sheet. These are consistent with magnetic discontinuities separating the denser and colder ambient plasma from the reconnected plasma inside the jets \citep{Zhou2009}. 
In conjunction with the two jets (shaded regions) we observe hotter plasma, as the count level for FPI is lower while the plasma is mainly detected by FEEPS at higher energies $\sim 100$ keV, especially in Jet 2. For these reasons, the moments from FPI are not always fully representative in jet regions, as suggested also by the discrepancy between the particle density measured by FPI and that obtained from the spacecraft potential, as shown in Figure \ref{Fig2}(c). In the region between the jets, where $B_z\sim |\textbf{B}|$ has a local minimum, the plasma is denser in correspondence of lower energies with respect to the plasma in the jets, and it is more similar to that in the plasma sheet. However, the difference in density is not large and this plasma is probably a mixture of ambient and reconnected plasma.
    
Figure \ref{Fig2}(d) shows the ion velocity components, obtained from FPI-DIS data, and panel \ref{Fig2}(e) reports the $x$ and $y$ ion velocity perpendicular components together with the $\textbf{E}\times \textbf{B}/B^2$ velocity (evaluated only when $\textbf{B} > 0.3$ nT) along $x$ and $y$. 
The differences between the $\textbf{E}\times \textbf{B}/B^2$ velocity and that measured by FPI-DIS in the reconnection jets are due to the part of thermal plasma in the jets lying beyond the FPI energy range. Both jets propagate earthward with $V_{x}>0$. In the first jet $V_{x}$ is larger, while for the second jet the $V_{y}$ dominates. Also, from the profiles in panel \ref{Fig2}(e), Jet 2 is characterized by dominant $y$ velocity, larger than the $x$ component in Jet 1. The standard timing analysis \citep{issi98} is applied on $B_z$ at 06:45:29 UT (Jet 1) and 06:46:12 UT (Jet 2) to evaluate the propagation speed of each jet front. For Jet 1, we obtained a jet's velocity of $\sim$ 330 km/s mainly along the $x$ direction. For Jet 2, the propagation speed is of $\sim$ 560 km/s directed in the $(x,y)$ plane, tilted of about 32 degree with respect to $x$. The velocity magnitudes are comparable to the local Alfv\'{e}n speed $V_A \sim$ 600 km/s. Thus the observations and the timing analysis results, suggest that the first, slower jet, propagating earthward, could act as an obstacle for the second, faster jet that could move toward the $y$ direction.

 \section{Physical interpretation}
 
\begin{figure}
        \centering
        \includegraphics[width=0.69\textwidth]{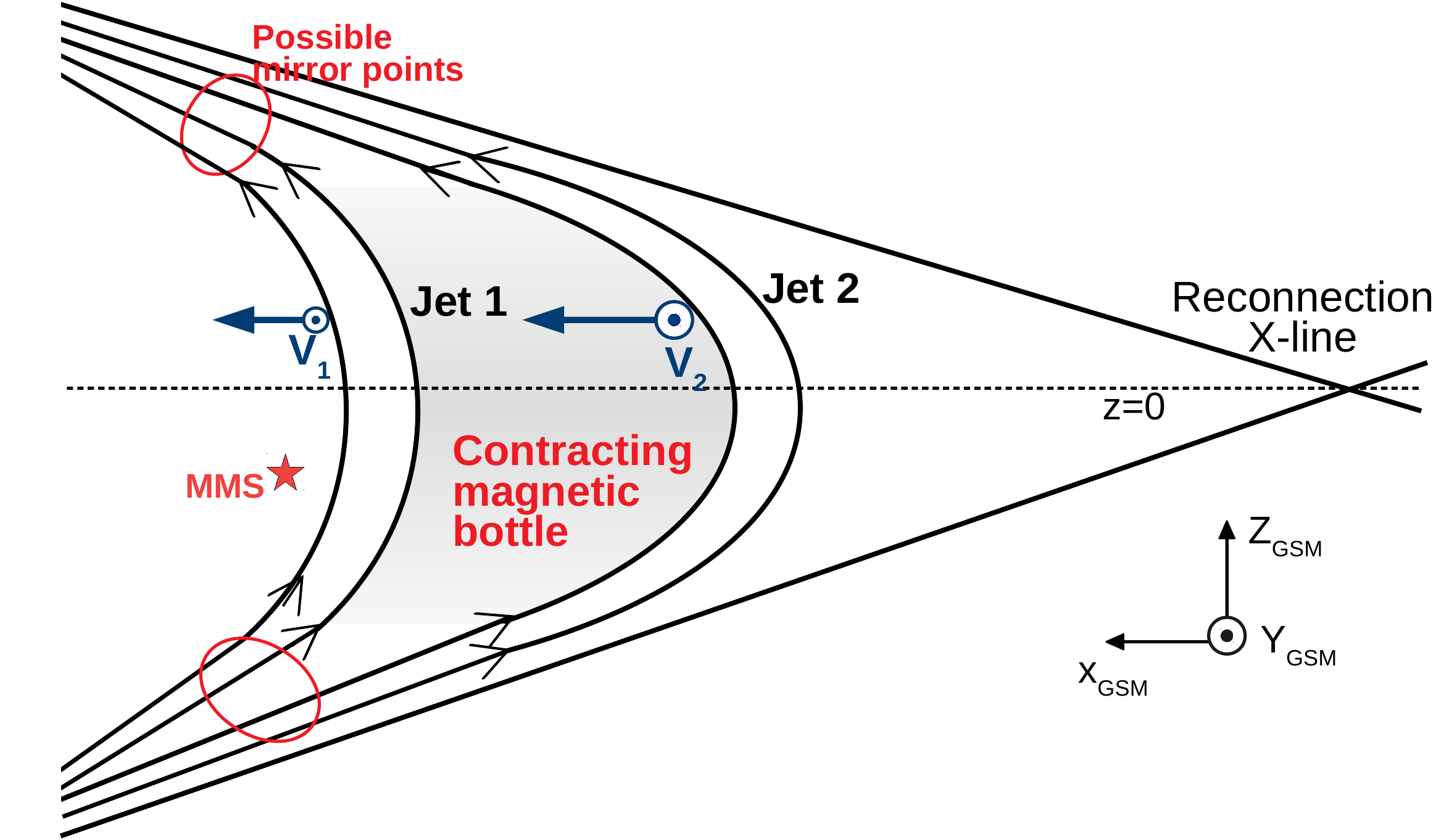}
        \caption{Sketch of the possible scenario. Jet 1 is slower and propagates mainly along the $x$ direction. Jet 1 could act as an obstacle for Jet 2, which is faster and deviates toward the $y$ direction. The jet interaction forms a magnetic bottle that contracts, due to the different jet propagation speeds.}\label{Fig4}
    \end{figure}

The $B_z$ and $E_y$ field profiles, similar to those reported in Figure \ref{Fig2}(a) and (b), have been observed in Particle-In-Cell simulations of unsteady reconnection where the dissipation at the jets was dominated by ions \citep{Sitnov2009}. The topology observed in these simulations (Figure 2 and 3 in \citet{Sitnov2009}) suggests that the observed regions of very strong $B_z$ and $E_y$ fields are consistent with sharp fronts propagating with the Alfv\'{e}n speed. This scenario is consistent with the observations in Figure \ref{Fig2}, suggesting the passage of two Alfvenic jet fronts. 
Also, in a classical two-dimensional scenario we would expect to observe earthward-directed plasma jets, which are mostly along the GSM X direction and whose directions correspond to the jet front propagation direction \citep{Runov2009,Runov2011}. However, in case of multiple jet fronts, the plasma in the second jet can be deflected by the first jet and start to move along the Y direction, as reported both in numerical simulation \citep{Ge2011} and in-situ observations \citep{Zhou2009}. This is in agreement with the increase of $V_y$ in correspondence of the Jet 2 in Figure \ref{Fig2}(d) and (e). Therefore, the scenario with two consecutive jets could explain the MMS observations reported in Figure \ref{Fig2}.
A schematic cartoon of this possible scenario is shown in Figure \ref{Fig4}, similarly to the one obtained in a numerical simulation \citep{Birn2011}, suggesting the formation of a magnetic bottle between the jets. 

\begin{figure}
        \centering
        \includegraphics[width=1.1\textwidth]{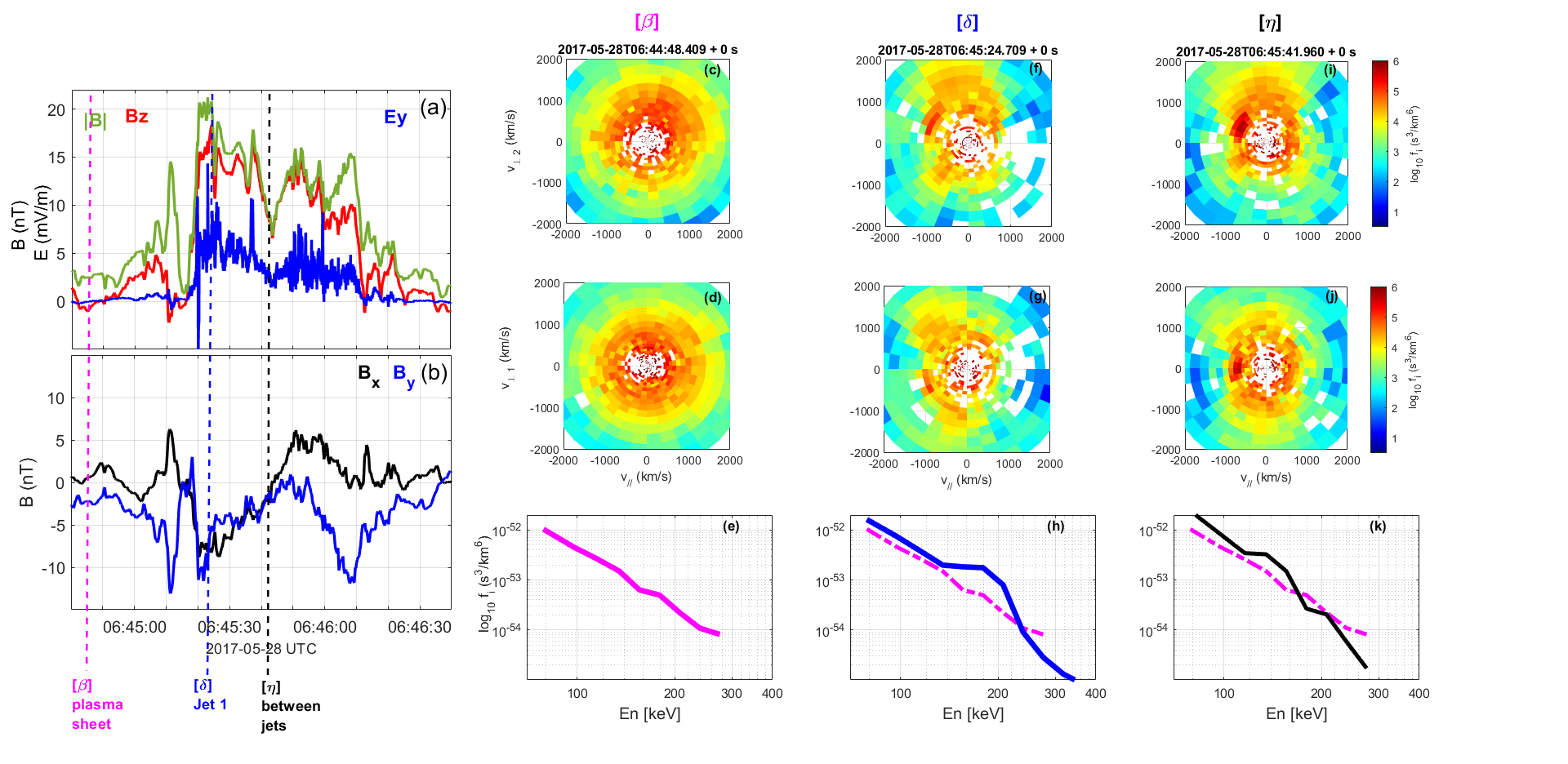}
        \caption{Panel (a) shows the $B_z$ magnetic field (red line), total magnetic field (green line) and $E_y$ electric field (blue line). Panel (b) shows the $x$ (black) and $y$ (blue) components of the magnetic field. Right panels show the ion distribution functions, detected by FPI, in the $(v_{\perp,2}, v_{//})$ and $(v_{\perp,1}, v_{//})$ planes at the cut $[\beta]$, $[\delta]$ and $[\eta]$. $\textbf{v}_{//}$ is along the local magnetic field, $\textbf{v}_{\perp,1}=\textbf{v}\times\textbf{b}$, $\textbf{v}_{\perp,2}=\textbf{v}\times\textbf{v}_{\perp,1}$, where $\textbf{v}=\textbf{V}_i/|\textbf{V}_i|$ and $\textbf{b}=\textbf{B}/|\textbf{B}|$. Bottom panel show the FEEPS energy spectra in $[\beta]$ (magenta line, averaged between 06:44:48 and 06:44:51 UT), $[\delta]$ (blue line), and $[\eta]$ (black line). Ion distributions in the same regions (or cuts) are measured at the same instant of time reported at the top of each column.}\label{Fig5}
    \end{figure}

\begin{figure}
        \centering
        \includegraphics[width=0.85\textwidth]{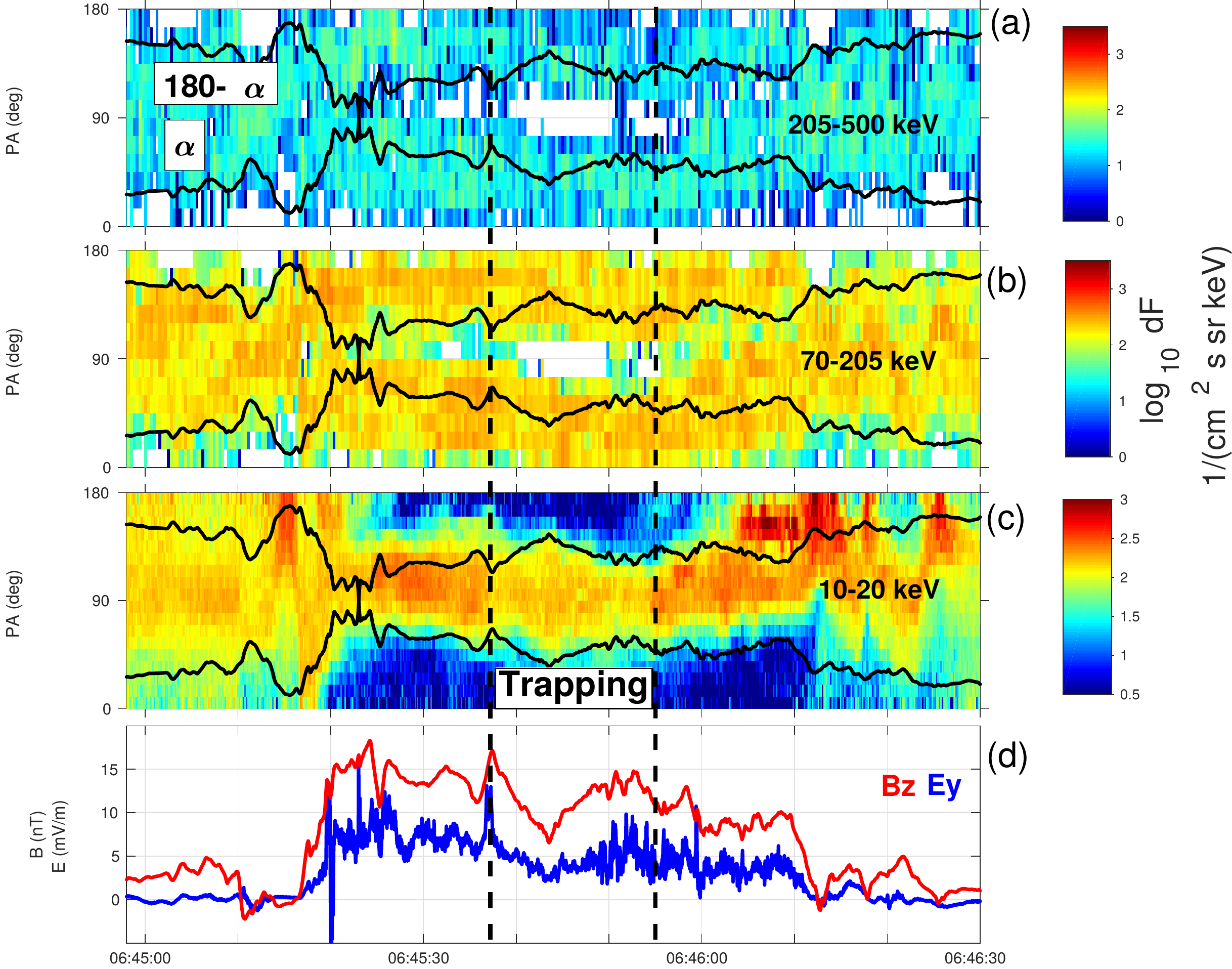}
        \caption{Ion pitch angle distributions in the energy range of [205-500] keV (a), [70-205] keV (b) and [10-20] keV (c). Panels (a) and (b) are obtained from FEEPS measurements, while panel (c) from FPI measurements. Black lines represent the local value of the trapping angles $\alpha$ and $180-\alpha$. Panel (d) shows $B_z$ (red line) and $E_y$ (blue line) profiles.}\label{Fig6}
    \end{figure}

Figure \ref{Fig5} shows the ion distribution functions in different regions. Panel (a) shows the $B_z$ and $|\textbf{B}|$ profiles, almost matching each other, together with $E_y$. Panel (b) shows the magnetic field components $B_x$ and $B_y$. The vertical lines ($[\beta]$, $[\delta]$ and $[\eta]$) in Figure \ref{Fig5}, denote the regions where the ion distribution functions are studied.
The cut $[\beta]$ is taken in the plasmasheet before the front passes, where $|\textbf{B}| \sim$ 3 nT and $B_x \sim$ 0 nT. Panels (c) and (d) show FPI ion distribution functions in $[\beta]$, projected in different planes (see caption of Figure \ref{Fig5}). Panel (e) displays the ion energy spectra detected by FEEPS in the region $[\beta]$ (magenta line), which is averaged between 06:44:48 and 06:44:51 UT to consider the main plasma sheet conditions. By fitting that spectra with a power law $f(E_i) \propto E_i^{\gamma}$, we obtain the spectral index $\gamma=-4$, similar to those reported in \citep{Haaland2010} for ions in the Earth's magnetotail. In the region $[\eta]$ between the jets, close to the CS center, the ion distribution function detected by FPI is less isotropic with respect to the one in the plasmasheet, as shown in Figure \ref{Fig5}(i) and (j). In this region most of the ions measured by FPI are distributed along the perpendicular direction. This is also visible in Figure \ref{Fig6}(c), showing perpendicular pitch angle distribution between the two jets, for ions with energies between $10-20$ keV.
Panel (k) compares the FEEPS ion energy spectra in the plasmasheet (magenta line), and between the jets (black line). 
The comparison shows that at all energies up to $150$ keV ($\sim 10-15$ times the thermal energy $\sim 10$ keV) the flux between the jets (black line) is higher with respect to the ambient plasma sheet population (magenta line).
This suggests that, in the region between the jest, ion energization is ongoing.
The ion distribution functions in panels (f), (g), and (h), correspond to the region inside the first jet denoted with the cut $[\delta]$ (blue vertical line) in Figure \ref{Fig5}. In this region the ion distribution measured by FPI is strongly anisotropic (panels (f) and (g)). 
By comparing the energy spectra detected by FEEPS (panel (h)) we can observe that the shape of the spectra inside Jet 1 (blue line) is different from the one of the ambient population (magenta line).  This spectrum is also different from the spectrum observed in between the jets (black line in Figure \ref{Fig5} (k)). The flux inside the first jet starts to increase over the ambient flux from $\sim 150$ keV and shows a plateau until  $\sim 200$ keV, while at energy below $150$ keV the flux is comparable to the ambient population (magenta line). Ions observed within the Jet 1 are likely ions that have been earlier accelerated at the reconnection site and are then transported inside the jet.

To understand the possible acceleration mechanism acting between the two jets, we study the pitch angle distributions (PADs). Figure \ref{Fig6} shows the PADs in three energy ranges: [205-500] keV (a), [70-205] keV (b), and [10-20] keV (c). 
As discussed above, the behaviour of the magnetic field suggests the formation of a magnetic bottle between the jets. We suggest that the bottle has mirror points along the magnetic flux tube, close to the CS rather than at the magnetic foot points, as also shown in \citep{Borissov16}. Possible mirror points are shown in the sketch in Figure \ref{Fig4}, and are formed locally due to the magnetic field transient dipolarization caused by the first jet (see Figure 3 in \citet{Birn2011}). We argue that ions from the thermal plasma distribution at $\sim 10$ keV between the two jets are magnetized and trapped inside the bottle. As the bottle contracts due to the different jet propagation speeds (the second being faster), ions are accelerated along the magnetic field due to the adiabatic conservation of the longitudinal moment $J=\oint p_{\parallel} dl= \rm const$.
We can estimate the loss cone angle $\alpha=\sin^{-1}\left(\sqrt{{\mid \textbf{B} \mid}/{B_{max}}}\right)$, where $B_{max} = $ 20 nT is the maximum value of the magnetic field magnitude in the whole interval reported in Figure \ref{Fig2}, and $|\textbf{B}|$ is value of the local magnetic field magnitude. The solid black lines in Figure \ref{Fig6} represent the loss cone angles $\alpha$ and $180^{\circ}-\alpha$. Ions having PA between the black lines are locally trapped in the bottle, while ions having PA below and above the black lines are not-trapped.

The PADs in between the jets, that is within the magnetic bottle (region marked by vertical lines in Figure \ref{Fig6}), have different features depending on the energy range. At thermal energies $10-20$ keV the pitch angle (PA) is mostly perpendicular, also in agreement with the distributions shown in Figure \ref{Fig5}(i) and (j). Higher energy ions $70-205$ keV, panel (b), show a more field-aligned distribution between the jets, with a part of the distribution populating the loss cone. This effect is continuing further at even higher energies ($205-500$ keV), panel (a).
The energy spectra reported Figure \ref{Fig5}(k) (showing an increase in the ion flux up to $\sim 150$ keV) and the PADs in Figure \ref{Fig6}, strongly suggest that plasma sheet ions having energies from $\sim 10-20$ keV are trapped between the jets and start to be accelerated by the contracting magnetic bottle via a Fermi-like mechanism. The energy range $70-150$ keV includes ions which are still trapped and are accelerated by the Fermi mechanism, although they are likely undergoing very few bouncing along shrinking magnetic field lines. On the other hand, ions having energies larger than $150$ keV start to fall into the loss-cone, be un-trapped, and are no more accelerated. 
It should be noted that the energy ranges used for the panels (b) and (c) in Figure \ref{Fig6} have been chosen in order to have a sufficient number of counts to best illustrate this effect and it is not possible to have a finer energy distribution. It should be also noted that no measurements unfortunately exist between $30$ keV and $70$ keV due the energy gap between FPI and FEEPS.
Only a few particles with PA $\sim 90^{\circ}$ and higher energy are observed in Figure \ref{Fig6}(a) and (b). This could also be due to the fact that FEEPS cannot distinguish between protons and heavier ions, the last being predominant above $150$ keV in multiply charged state (as shown by \cite{Allen16, Cohen2017}). For a fixed ion mass, the energy needed to escape the bottle is proportional to $q_i^2$, where $q_i$ is the ion charge. Therefore, the ions inside the black lines in Figure \ref{Fig6}(a) and (b) at energies $> 150$ keV could be multiply charged heavy ions still trapped inside the magnetic bottle. Unfortunately, it was not possible to study the thermal ion composition because data from the Hot Plasma Composition Analyzer (HPCA) instrument \citep{Young2016} are not available for the investigated time period. Also, we note that the vertical dashed lines in Figure \ref{Fig6} mark the region where we observe perpendicular PAD, almost coinciding with the lines denoting the region between the jets in Figure \ref{Fig2}. 

Betatron acceleration in the region between the jets could also be at work \citep{Zharkova2011}, although probably it is not the dominant effect
because the magnetic field intensity exhibits local variations and yet there is not a strong increasing trend of $|\textbf{B}|$.
Furthermore, considering the magnetic field variation observed between the jets, the ions could be accelerated via betatron mechanism reaching $\sim 7$ times their thermal energy. These energies are below those observed at $\sim 10-15$ times the thermal energy. Also other scenarios can be considered to interpret the observations, as for example the passage of a well structured single jet. In this case the observed energetic ions would be accelerated by magnetic reconnection. But this scenario is inconsistent with the observed PAD distribution, because plasma accelerated via magnetic reconnection is generally associated to an isotropic distribution \citep{Hoshino2001, Oieroset2002}. Also, considering a single jet passage will be difficult to explain the increase of $V_y$ observed in Figure \ref{Fig2}. Thus, it is more consistent to interpret the results following the scenario of MMS observing ion acceleration between two consecutive reconnection jets.

	\section{Conclusions}
In this paper we discuss high resolution MMS spacecraft measurements of earthward propagating plasma jet fronts as drivers for suprathermal ion acceleration during unsteady magnetic reconnection. The observed magnetic and electric field profiles as well as plasma measurements are consistent with the formation of a magnetic bottle between the two jets. Since the fronts have different propagation speeds, the second jet being faster, the bottle contracts. We propose that thermal ions from the ambient plasma population, having perpendicular pitch angle distribution, are initially trapped and, as the bottle contracts, are gradually energized along the parallel direction through a Fermi-like mechanism until they fall into the loss cone. This mechanism accelerates the ions up to $\sim 150$ kev, which is almost $\sim 10-15$ times the thermal energy. The observed jet fronts are reminiscent of the downward-moving reconnected field lines in collapsing magnetic traps during solar flares, where strong particle acceleration occurs. In contrast to the classical scenario of magnetic traps where mirror points are formed at foot points, our observations suggest that mirror points could instead be formed closer to the CS due to the interaction of subsequents jets, as also suggested in \citep{Borissov16}. We speculate that during unsteady reconnection many magnetic bottles could form upon interaction of jets, which could lead even to acceleration to high energies.  
This ion acceleration mechanism can have far-reaching implications for many astrophysical environments, in particular solar and stellar flares where unsteady reconnection and jet fronts are ubiquitous.

\section{Acknowledgments}
MMS data can be retrieved from the MMS Science Data Center at https://lasp.colorado.edu/mms/sdc/. 
F.C. is grateful to the LPP team for their support. S. P. has been supported by the Agenzia Spaziale Italiana under the contract ASI-INAF 2015-039-R.O “Missione M4 di ESA: Partecipazione Italiana alla fase di assessment della missione THOR”. Research at Southwest Research Institute was founded by NASA contact NNGO4EB99C.

\bibliography{Bibliography}

\section{Figures}

\end{document}